\RequirePackage{ifpdf}
\documentclass[11pt]{JHEP3}

\pdfoutput=1

\usepackage{amssymb}
\usepackage{graphics}
\usepackage{epsfig}
\usepackage{graphicx}
\usepackage{subfigure}
\usepackage{bm}

\newcommand{\be}{\begin{equation}}
\newcommand{\ee}{\end{equation}}
\def\bea{\begin{eqnarray}}
\def\eea{\end{eqnarray}}

\newcommand{\Mp}{M_{\rm p}}

\newcommand{\Mpc}{\,{\rm Mpc}}
\newcommand{\Kpc}{\,{\rm Kpc}}

\newcommand{\z}{\zeta}

\newcommand{\e}{\epsilon}
\newcommand{\g}{\gamma}
\newcommand{\G}{\Gamma}
\newcommand{\Dn}{\delta N}
\newcommand{\prm}{\prime}

\newcommand{\s}{\sigma}
\newcommand{\mphi}{m_\phi^2}
\newcommand{\msig}{m_\s^2}
\newcommand{\vp}{\varphi}
\newcommand{\dtvp}{\dot{\vp}}

\newcommand{\sosc}{\s_{\rm osc}}
\newcommand{\rdec}{r_{\rm dec}}
\newcommand{\fnl}{f_{\rm NL}}

\newcommand{\nz}{n_\z}
\newcommand{\nfnl}{n_{\fnl}}
\newcommand{\gfnl}{\gamma_{\fnl}}
\newcommand{\gnl}{g_{\rm NL}}

\def\bk{{\bf k}}

\title{Scale-dependent non--Gaussianity and the CMB Power Asymmetry}

\author{Christian T.~Byrnes,
Ewan R. M. Tarrant \\
Astronomy Centre, University of Sussex, Falmer, Brighton BN1 9QH, UK}

\abstract{We introduce an alternative parametrisation for the scale dependence of the non--linearity parameter $\fnl$ in quasi--local models of non--Gaussianity. Our parametrisation remains valid when $\fnl$ changes sign, unlike the commonly adopted power law ansatz $\fnl(k)\propto k^{\nfnl}$. We motivate our alternative parametrisation by appealing to the self--interacting curvaton scenario, and as an application, we apply it to the CMB power asymmetry. Explaining the power asymmetry requires a strongly scale dependent non--Gaussianity. We show that regimes of model parameter space where $\fnl$ is strongly scale dependent are typically associated with a large $\gnl$ and quadrupolar power asymmetry, which can be ruled out by existing observational constraints.\\} 

\keywords{scale dependence, curvaton, non-Gaussianities, self-interactions, bispectrum, inflation, power asymmetry}

\begin{document}

%-------------------------------------------------------------------------------------------------------------
\section{Introduction}\label{sec:intro}

The paradigm of inflation~\cite{Starobinsky:1979ty,Starobinsky:1980te,Kazanas:1980tx,Sato:1980yn,Guth:1980zm} is the simplest framework which explains the origin of the primordial cosmological fluctuations~\cite{Mukhanov:1981xt,Hawking:1982cz,Guth:1982ec,Starobinsky:1982ee,Bardeen:1983qw}. The detailed information which is encoded in these fluctuations over a range of cosmological scales offers the opportunity to test the plethora of inflationary models which have been proposed in the literature. In particular, the local ansatz for non--Gaussianity has become a popular description for modelling deviations from a purely Gaussian distribution of primordial perturbations.

For the simplest model where the primordial perturbation is a quadratic local function of a single Gaussian field, non--Gaussianity is usually parametrised in terms of a single constant parameter, $\fnl$, which measures the amplitude of the bispectrum normalised to the square of the power spectrum of primordial curvature fluctuations~\cite{Komatsu:2001rj}.  In this case $\fnl$ is scale--independent by construction.  In general however $\fnl$ is not constant over all scales, as expected from theoretical considerations~\cite{Byrnes:2008zy,Byrnes:2009pe,Byrnes:2010xd,Huang:2010cy,Kobayashi:2012ba,Dias:2013rla} (see~\cite{Chen:2005fe} for considerations of scale dependent equilateral non--Gaussianity). In this paper we exclusively consider non--Gaussianity which reduces to the local shape in the limit of scale invariance.  Scale--dependence can arise due to multiple fields contributing to the curvature perturbation, a curved field space metric \cite{Byrnes:2012sc,Gao:2014fva}, or due to non--linear evolution of modes after they leave the Hubble radius during inflation. This scale dependence is typically parametrised as a power law, 
\bea \fnl(k)=\fnl^0(k/k_0)^{\nfnl}, \eea
where $\fnl^0$ is the amplitude of $\fnl$ measured at some pivot scale $k_0$ and $\nfnl$ is a constant (see, e.g.,~\cite{Sefusatti:2009xu,Byrnes:2009pe,Becker:2012je} for details). Re--writing this as
\be
\nfnl\equiv\frac{{\rm d}\,{\rm ln}|\fnl|}{{\rm d}\,{\rm ln}k}\,,
\label{eq:nfnl_std}
\ee
one can clearly see that $\nfnl$ diverges whenever $\fnl$ changes sign. The self--interacting curvaton model~\cite{Enqvist:2005pg,Byrnes:2011gh} and the dynamics of perturbative reheating in the axion--like model studied in Ref.~\cite{Meyers:2013gua} are two examples where $\fnl$ can change sign over the observable range of scales, $10^{3}\Kpc\lesssim k^{-1}\lesssim10^{4}\Mpc$.  In such examples the scale--dependence of $\fnl$ cannot be described by a simple power law over scales which include $\fnl$ close to zero. We define $\fnl$ as the ratio of the bispectrum to the square of the power spectrum in the usual way, which for an equilateral configuration reduces to
\bea \fnl(k)=\frac{5}{18}\frac{B_{\zeta}(k,k,k)}{P_\zeta^2(k)}. \eea 
In this paper we introduce an alternative parametrisation for the scale dependence of $\fnl(k)$: 
\be
\fnl(k)=\fnl^0+\gfnl\,{\rm ln}\,\left(\frac{k}{k_0}\right)+\cdots\,,
\label{eq:fnlk}
\ee
where $\gfnl$ is a constant, and $\fnl^0$ is the value of $\fnl$ measured at the pivot scale $k_0$ for an equilateral configuration. Unlike the conventional parametrisation $\fnl\propto k^{\nfnl}$, ours remains valid when $\fnl$ vanishes, and is theoretically well--motivated as we demonstrate by appealing to the self--interacting curvaton model as an explicit example where $\fnl(k)$ can exhibit a sign change over an observable range of scales. We also show that even in cases with $|\gfnl|\gg1$, it is often a good approximation to take this parameter as a constant over the observable range of scales.\\

As an important application of our results, we study the observed dipolar power asymmetry of the primordial power spectrum \cite{Eriksen:2003db}. On large scales this is observed to be a large effect, but the asymmetry must be at least an order of magnitude smaller on small scales, where the effect is not observed but tightly constrained \cite{Akrami:2014eta}. If not a statistical fluctuation, the asymmetry may arise through the non--Gaussian coupling between a large amplitude super horizon mode and the modes where the asymmetry is observed. The amplitude of the non--Gaussianity must also vary by an order of magnitude in order to explain the reduction of this effect on small scales. Such a large variation cannot be accurately described by the conventional parametrisation $\fnl(k)=\fnl^0(k/k_0)^{\nfnl}$ with constant $\nfnl$, which is only valid under the assumption $|\nfnl\ln(k/k_0)|\ll1$, however we explicitly show that it can be accurately described by Eq.~(\ref{eq:fnlk}). 

We also show that regions of model parameter space where $\fnl$ varies sufficiently quickly with scale and is large on the scales where the asymmetry is large typically give rise to a large $\gnl$ and quadrupolar modulation of the power spectrum. Unlike $\fnl$ and the quadrupolar asymmetry, these two additional effects are close to scale invariant and for typical parameter values too large to be consistent with existing observational constraints.

In the next section we introduce our parametrisation of strongly scale dependent $\fnl$ and apply it to the self--interacting curvaton scenario in Sec.~\ref{sec:selfintcurv}, deriving new analytic solutions in Sec.~\ref{sec:analytic_sol} which are valid in the limit that the quadratic term of the curvaton potential dominates. Unlike previous analytic solutions, ours  takes into account the evolution of the curvaton during inflation and reheating, and we show that this evolution has a large effect which may give rise to qualitatively new behaviour of the non--Gaussianity. We apply our formalism to the power asymmetry in Sec.~\ref{sec:asymm} and study the difficulty in keeping $\gnl$ and the quadrupolar power asymmetry small in Sec.~\ref{sec:g}. Finally we conclude in Sec.~\ref{sec:conc}. Throughout this paper we work in units where $\hbar=c=1$, and we set the reduced Planck mass $\Mp=1$.
%-------------------------------------------------------------------------------------------------------------
\section{Strongly scale dependent non--Gaussianity}\label{sec:non-g}

During inflation, it is normally accurate to parametrise the power spectrum by a power law, both from a theoretical and observational perspective. For both single and multi--field inflation, it is normally accurate to treat the spectral index as a constant over a much larger range of scales than we can observe (unless the running is large). The spectral index is related to the slow--roll parameters, which are normally required to be small in order for inflation to last many efoldings. A spectral index of local $\fnl$, $\nfnl$, was introduced in~\cite{LoVerde:2007ri,Byrnes:2008zy,Sefusatti:2009xu,Byrnes:2009pe,Becker:2012je} in analogy with the power spectrum and its value is also proportional to the slow--roll parameters.  However, in contrast to the power spectrum, the scale dependence of $\fnl$ can be greatly enhanced by numerical factors much greater than unity \cite{Byrnes:2011gh}. In such cases, it is not accurate to treat $\nfnl$ as a constant, but it is precisely such cases which are the most observationally interesting because the current tight constraints on $\fnl$ mean we can only hope to measure both $\fnl$ and its scale dependence if the scale dependence is much larger than the observed scale dependence of the power spectrum, for forecasts see e.g.~\cite{Sefusatti:2009xu,Shandera:2010ei,Giannantonio:2011ya,Raccanelli:2014awa,Leistedt:2014zqa}. 

Motivated by the observation that regions of large $\nfnl$ tend to correspond to regimes where $\fnl$ is small, we introduce the new parametrisation Eq.~(\ref{eq:fnlk}) which remains valid as $\fnl$ crosses zero. Provided that $\fnl$ grows to large values within the scales that we can probe, this case is observationally interesting. As an additional motivation, the observed power asymmetry which we discuss in Sec.~\ref{sec:asymm}~ requires a strongly scale dependent $\fnl$, with $\nfnl\sim-0.5$. For most models, treating $\nfnl$ as constant will only be valid for about $2$ $e$--foldings when its magnitude is so large, which is a much smaller range of scales than over which the observation is made. Our parametrisation can be easily related to the standard one
\bea 
\gfnl=\frac{d \fnl(k)}{d\ln k}=\fnl(k)\nfnl(k)\,, \label{relate}
\eea
but our definition remains valid in the limit $\fnl^{0}=0$. Furthermore we will see that our parametrisation remains valid (for constant $\gfnl$) over a large range of scales in the observationally interesting regime, which includes the requirement that $|\gfnl|\gg1$. In contrast, in such regimes where $\fnl$ is strongly scale dependent, it is clear from (\ref{relate})  that $\nfnl$ must be equally strongly scale dependent in order for $\gfnl$ to be a constant.

%-------------------------------------------------------------------------------------------------------------
\subsection{The self--interacting curvaton scenario}\label{sec:selfintcurv}

In this section we motivate the parametrisation Eq.~(\ref{eq:fnlk}) by studying the scale dependence of $\fnl$ in the self--interacting curvaton scenario. The model is described by the potential
\be
W(\phi,\s)=V(\phi) + \frac12\msig\s^2 + \lambda\s^n\,,
\label{eq:potential}
\ee
where $\s$ is the curvaton field, and inflation is driven by $V(\phi)$. We will consider $n\ge4$, with $n$ even. The phenomenology of this model has been extensively studied in previous work, see e.g.~\cite{Dimopoulos:2003ss,Enqvist:2005pg,Enqvist:2008gk,Huang:2008bg,Byrnes:2010xd}, where it was found that observable quantities (such as $\fnl$) can depart substantially from the standard quadratic curvaton prediction due to self--interactions. It is useful to define the following quantity
\be
s_k\equiv s(t_k)=\frac{2\lambda\s^{n-2}(t_k)}{\msig}\,,
\label{eq:sk}
\ee
which gauges the strength of the curvaton self interaction relative to the mass term in the potential. Throughout this paper, we denote the time at which the mode $k$ left the Hubble radius by $t_k$. For the sake of brevity we will often write $X_k=X(t_k)$ for scale dependent quantities which are to be evaluated at the epoch of Hubble exit.

After inflation has ended $\phi$ begins to oscillate about the minimum of its potential, decaying to radiation as it does so, which comes to dominate the energy density of the universe. We employ the sudden decay approximation to model the decay of the curvaton field, which assumes that the curvaton decays instantly into radiation when $H=\G_\s$. While the sudden decay approximation does not properly treat the gradual transition from an oscillating scalar field to a radiation bath, the behaviour far from this transition period, and the resulting ratios of energy densities are well captured by the approximation. As has been shown in previous work~\cite{Sasaki:2006kq,Meyers:2013gua} this is enough to quite accurately reproduce the effects of the decay process on primordial observables. If the curvaton is coupled directly to the inflaton or to other scalar fields, then it can decay non--perturbatively through parametric resonance~\cite{Traschen:1990sw,Kofman:1994rk,Enqvist:2008be,Sainio:2012rp}. Since in the potential Eq.~(\ref{eq:potential}) the fields are coupled only through gravity, we do not consider this possibility here. 
Sufficiently deep into the phase of coherent oscillations of the curvaton field, but before it has decayed into radiation, the curvaton energy density is very well described by the quadratic contribution, $\rho_\s=\msig\sosc^2/2$, where $\sosc=\sosc(\s(t_k))$ is the amplitude of the curvaton oscillations about the quadratic part of the potential. We assume that the curvature perturbation $\z$ is generated solely by fluctuations of the curvaton field and we neglect the inflaton fluctuations.  Appealing to the $\Dn$ formalism~\cite{Sasaki:1995aw,Sasaki:1998ug,Lyth:2005fi}, the curvature perturbation can be written at third order as (see e.g.~\cite{Sasaki:2006kq})
\bea
\z_\bk &=& N^\prm(t_k)\delta\s_\bk(t_k)+\frac12N^{\prm\prm}(t_k)(\delta\s\star\delta\s)_\bk(t_k)  +\frac16N^{\prm\prm\prm}(t_k)(\delta\s\star\delta\s\star\delta\s)_\bk(t_k) + \dots \nonumber \\ 
&=& \frac{2\rdec}{3}\frac{\sosc^\prm}{\sosc}\delta\s_\bk(t_k)+\frac{\rdec}{3}\left[ \frac{\sosc^{\prm\prm}}{\sosc} + \left(\frac{\sosc^\prm}{\sosc} \right)^2\right](\delta\s\star\delta\s)_\bk(t_k) \nonumber \\
&+& \frac{\rdec}{9}\left[ \frac{\sosc^{\prm\prm\prm}}{\sosc} +3\frac{\sosc^{\prm\prm}\sosc^\prm}{\sosc^2} \right](\delta\s\star\delta\s\star\delta\s)_\bk(t_k) + \dots\,,
\eea
where 
\be
\rdec=\frac{3\rho_\s}{3\rho_\s+4\rho_\g}\Big|_{\rm dec}\,.
\ee
Subscript `${\rm dec}$' denotes the time of curvaton decay, and $(^\prm)\equiv\partial_{\s_k}$. The convolutions are defined by $(\delta\s\star\delta\s)_\bk(t_k) =(2\pi)^{-3}\int {\rm d}{\bf q}\delta\s_{\bf q}(t_k)\delta\s_{\bk - {\bf q}}(t_k)$.  $N(t_k)$ denotes the number of $e$--foldings from an initial spatially--flat hypersurface at $t_k$, to a final uniform density surface after the decay of the curvaton. The non--linearity parameters $\fnl(k)$ and $\gnl(k)$ are given by (see e.g.~\cite{Sasaki:2006kq})
\bea
\fnl(k)&=&\frac54\frac{f_{\rm osc}}{\rdec} -\frac53 -\frac56\rdec\,, \nonumber \\
\frac{54}{25}\gnl(k)&=&\frac94\frac{g_{\rm osc}}{\rdec^2}-9\frac{f_{\rm osc}}{\rdec}-\frac92\left(f_{\rm osc}-\frac{10}{9}\right)+10\rdec+3\rdec^2\,, \nonumber \\
f_{\rm osc}&\equiv& 1 + \frac{\sosc\sosc^{\prm\prm}}{\sosc^{\prm2}}\,, \qquad
g_{\rm osc}\equiv\frac{\sosc^2\sosc^{\prm\prm\prm}}{\sosc^{\prm3}} +3\frac{\sosc^{\prm\prm}\sosc}{\sosc^{\prm2}}\,.
\label{eq:curvaton_fnl}
\eea
Scale dependence is manifest since $\sosc$ is a function of $\s_k$. If the curvaton potential were exactly quadratic, the equation of motion for $\s$ would be linear, and $f_{\rm osc}=1$, $g_{\rm osc}=0$. When the potential deviates from quadratic, derivatives of $\sosc$ contain information about the self interaction. As has been previously observed~\cite{Byrnes:2010xd,Enqvist:2005pg}, even a small correction to the quadratic potential can have an important effect on the level of non--Gaussianity while leaving the Gaussian perturbation essentially unchanged.  

In the limit $\rdec\ll1$, we see from Eq.~(\ref{eq:curvaton_fnl}) that $\fnl$ vanishes whenever $f_{\rm osc}=4\rdec/3\ll1$. Working in this limit and using the standard parametrisation for $\fnl(k)$, Eq.~(\ref{eq:nfnl_std}), the authors of Ref.~\cite{Byrnes:2011gh} studied the scale dependence of $\fnl$ (and $\gnl$). For $n\ge6$, $\fnl$ was found to be a highly oscillatory function of $\s_k$, resulting in a series of `spikes' in $\nfnl$ as $\fnl$ changes sign. At these points, the standard parametrisation cannot be used. 

%-------------------------------------------------------------------------------------------------------------
\subsection{Scale dependent $\fnl$ in the limit $s_k\ll1$}\label{sec:analytic_sol}

Analytic solutions in the limit of weak self--interactions have been obtained in previous studies~\cite{Enqvist:2005pg} however, these solutions are restricted to the regime where the curvaton remains frozen until the inflaton has decayed into radiation. The solution that we present below accounts for evolution of the curvaton field during inflation. One motivation for including the effects of curvaton dynamics during inflation is that these dynamics may be relevant for the CMB power asymmetry. Because we are interested in the parameter space with large non--Gaussianity, i.e.~$\rdec\ll1$, we assume throughout this paper that the curvaton does not effect the Hubble parameter. We have numerically checked that this is a good approximation.

In the limit $s_k\ll1$ the potential Eq.~(\ref{eq:potential}) is dominated by the mass term, and the effect of the curvaton self interaction is to weakly deform the potential away from pure quadratic form.  As the curvaton rolls down its potential, the magnitude of the self--interaction term relative to the mass term quickly decreases, until a point is reached where the potential is almost exactly quadratic. In this regime, the equation of motion for $\s$ is linear,
\be
\ddot\s+3H\dot\s+\msig\s\approx0\,, \qquad {\rm for} \quad t\ge t_q\,.
\label{eq:linearEoM}
\ee
The general solution can be expressed in the form $\s(t)=\sosc\left[ aY_1(t) + bY_2(t) \right]$ where $\sosc=\s(t=t_q)$, and $t_q$ is defined to be the time from which the linear equation above becomes valid.  If we make the additional assumption that $\s$ does not contribute to the Hubble rate, $Y_{1,2}(t)$ are functions of time which depend only on the inflationary background. Owing to the non--linearity of the curvaton equation of motion for $t<t_q$, $\sosc$ will be a non--linear function of $\s_k$. To calculate this functional dependence we assume that $\s$ remains slowly rolling until Eq.~(\ref{eq:linearEoM}) becomes a faithful description of the curvaton dynamics (note that if the curvaton is sufficiently light, it will remain in slow--roll even after inflation ends). This enables us to write $\sosc=\beta\s_{\rm SR}(t=t_q)$, for some constant $\beta$, where $\s_{\rm SR}$ is the solution obtained from the curvaton slow--roll equation of motion:
\be
\qquad 3H\dot\s + \msig\s + n\lambda\s^{n-1} \approx 0\,, \qquad {\rm for} \quad t\le t_q\,.
\ee
This equation can be directly integrated, to yield the solution at time $t=t_q$ for a generic inflationary background:
\be
\s_{\rm SR}(t_q)=\s_k\left[ \frac{e^{-\bar\eta_\s(n-2)\mathcal{I}(t_q)}}{1+\frac{n}{2}s_k[1-e^{-\bar\eta_\s(n-2)\mathcal{I}(t_q)}]} \right]^{\frac{1}{n-2}}\,, 
\label{eq:curvatonSR}
\ee
where
\be
\mathcal{I}(t_q)\equiv H^2_k \int^{t_q}_0\frac{{\rm d}t}{H(t)}\,,
\label{eq:Itq}
\ee
and the parameter $\bar\eta_\s\equiv\msig/3H_k^2$ is related to the curvaton slow-roll parameter
\be
\eta_\s(t_k) = \frac{W^{\prm\prm}_k}{3H^2_k}=\bar\eta_\s\left[ 1 + \frac12n(n-1)s_k \right]\,.
\label{eq:eta}
\ee
The curvaton does not contribute to the integral $\mathcal{I}(t_q)$, since we have neglected its backreaction to gravity. To a good approximation, the curvaton begins to oscillate when the Hubble rate becomes equal to the curvaton mass. An appropriate choice for $t_q$ is therefore $H^2(t_q)\sim\msig$. The integral $\mathcal{I}(t_q)$ is dominated by its upper limit, and so we have approximately $\mathcal{I}(t_q)\sim H^2_k/\msig\sim1/\bar\eta_\s$. Inserting this result into Eq.~(\ref{eq:curvatonSR}) and making the simplification $e^{-(n-2)}\sim0$ for $n\ge4$, we have:
\be
\sosc=\beta\s_{\rm SR}\approx\beta  e^{-\mathcal{O}(1)}\s_k\left( 1+\frac{n}{2}s_k  \right)^{-1/(n-2)}\,.
\label{eq:curvatonSR2}
\ee
Taking derivatives with respect to $\s_k$ and inserting into Eq.~(\ref{eq:curvaton_fnl}) we obtain the remarkably simple result for $\fnl$ in the limit $\rdec\ll1$:
\be
\fnl(k) = \frac{5}{4\rdec}\left[ 1-\frac12n(n-1)s_k  \right] -\frac53 \,.
\label{eq:fnl_analytic}
\ee
This expression relates $\fnl$ to the curvaton self--interaction strength $s_k$, which was defined in Eq.~(\ref{eq:sk}). The scale dependence of $\fnl$ is manifest since $s_k$ is a function of $k$. We compute this scale dependence explicitly in what follows. 
\FIGURE{
\epsfig{file=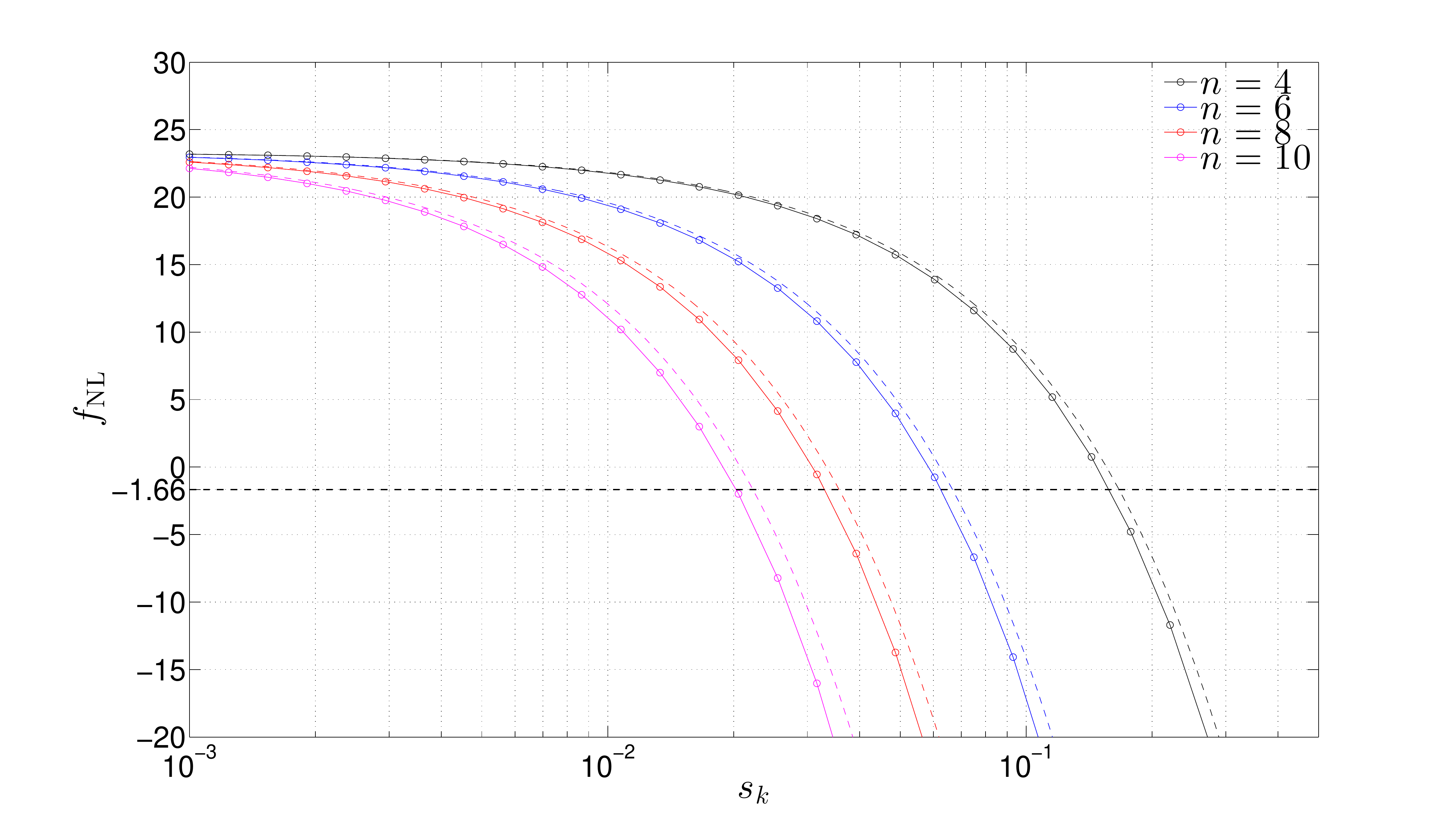,width=16cm} 
\caption{The asymptotic value of $\fnl$ (the final value which should be compared to observations) as a function of the self-interaction strength  $s_k$.  We have chosen an arbitrary value $\rdec=0.05$. Solid lines correspond to a numerical calculation, and dashed lines to the analytic result Eq.~(\ref{eq:fnl_analytic}). For the numerical calculation, we took $V(\phi)=\mphi\phi^2/2$, with $\phi_k=16$, $\bar\eta_\s=2.3\times10^{-2}$. For this and all other figures, we choose $\Gamma_\phi=H_{\rm osc}$, where $H_{\rm osc}$ is the Hubble rate at the time the inflation starts to oscillate. The intersection of the horizontal dashed line with the analytic solution indicate the points, $s_k=s_0$, where $\fnl=-5/3$. These values are given by Eq.~(\ref{eq:szero}).}
\label{fig:fNL_sk}
}
When $s_k=0$, we recover the standard quadratic curvaton result, $\fnl=5/(4\rdec)-5/3$, which is independent of $k$. By virtue of the approximation $e^{-\bar\eta_\s(n-2)\mathcal{I}(t_q)}\sim0$, our result for $\fnl$ has no dependence on $\bar\eta_\s$. Had we computed the integral $\mathcal{I}(t_q)$ exactly (which would require specifying the inflationary potential $V(\phi)$), $\fnl$ would develop a weak dependence on $\bar\eta_\s$. In the limit $s_k\ll1$ this dependence is negligible, and so to a very good approximation $\fnl$ is a function of $s_k$ only. To the best of our knowledge, this is the first time an analytic solution for $\fnl$ -- \emph{which is valid for arbitrary $\bar\eta_\s$} -- has been constructed for the self--interacting curvaton model.  An analytic solution for $\fnl$ in the limit $s_k\ll1$ was obtained in Ref.~\cite{Enqvist:2005pg}, and a solution for $n=4$ valid for all $s_k$ was presented in Ref.~\cite{Enqvist:2009zf}. Both of these solutions however are restricted to the regime where the curvaton remains frozen until the inflaton has decayed into radiation. To consistently realise this scenario within a typical inflationary background requires $\bar\eta_\s\ll10^{-7}$. As we shall see shortly, the scale dependence of $\fnl$ is directly proportional to $\bar\eta_\s$, and so to generate an interesting (large) scale dependence requires $\bar\eta_\s\gg10^{-7}$.

\FIGURE{
\epsfig{file=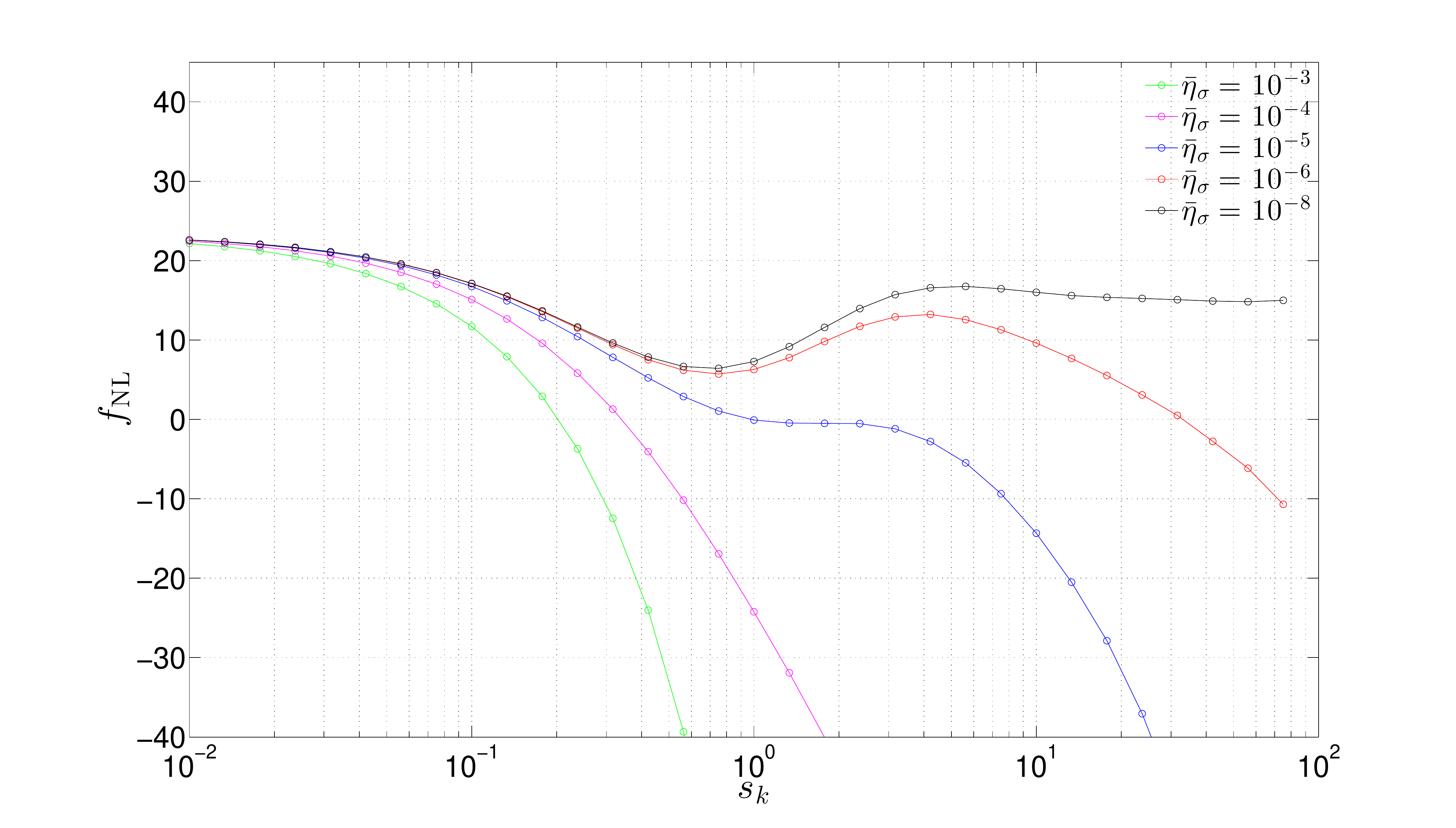,width=16cm} 
\caption{The asymptotic value of $\fnl$ as a function of the self interaction strength $s_k$,  for $n=4$, for various values of $\bar\eta_\s$. For large values of $\bar\eta_\s$, $\fnl$ can cross zero. For $\bar\eta_\s=10^{-8}$ (black line) the curvaton remains frozen until the inflaton has decayed, and we recover the results of Ref.~\cite{Byrnes:2011gh}. For $s_k\ll1$, $\fnl$ is independent of $\bar\eta_\s$, as revealed by the analytic solution Eq.~(\ref{eq:fnl_analytic}). All lines are computed numerically, with $V(\phi)=\mphi\phi^2/2$, $\phi_k=16$. We have chosen an arbitrary value $\rdec=0.05$.}
\label{fig:fs_n4}
}
In Fig.~\ref{fig:fNL_sk} we compare Eq.~(\ref{eq:fnl_analytic}) as a function of the self--interaction strength $s_k$, against a full numerical calculation, for various values of $n$. Our numerical technique (which is described in detail in Ref.~\cite{Meyers:2013gua}) computes the derivatives appearing in Eq.~(\ref{eq:curvaton_fnl}) for $\fnl$ exactly, accounting for gravitational backreaction of the curvaton. For illustration purposes, we choose a quadratic potential for the inflaton, $V(\phi)=\mphi\phi^2/2$, however we have confirmed that the results for $\fnl$ are largely insensitive to this choice. In our numerical calculations, the inflaton transfers its energy to radiation through a friction term $\G_\phi\dot\phi$, (i.e., $\ddot\phi+(3H+\G_\phi)\dot\phi+\mphi\phi=0$) which we introduce by hand into the equation of motion once $\phi$ is coherently oscillating. Throughout this paper, we set $\Gamma_\phi=H_{\rm osc}$, where $H_{\rm osc}$ is the value of the Hubble parameter at the onset of coherent oscillations of the inflaton field. With this value of $\Gamma_\phi$, the inflaton decays to radiation almost instantly. We have checked numerically that the results are largely insensitive to this choice.

It is possible for $\fnl$ to change sign for $n=4$ when the curvaton is allowed to roll during inflation, as clearly seen in Fig.~\ref{fig:fNL_sk}. This behaviour was not observed in earlier studies of the self--interacting curvaton model with $n=4$~\cite{Enqvist:2005pg,Byrnes:2011gh} which were restricted to the regime where the curvaton remains frozen throughout inflation. We do however numerically recover the results of~\cite{Byrnes:2011gh} for $n=4$ as $\bar\eta_\s$ is steadily decreased until the point where the curvaton remains frozen throughout inflation. This is illustrated in Fig.~\ref{fig:fs_n4}.

As can be seen from Fig.~\ref{fig:fNL_sk}, $\fnl$ changes sign as the self--interaction strength is varied. The standard parametrisation for $\fnl$, Eq.~(\ref{eq:nfnl_std}), is not valid if this sign changes occurs over the observable range of scales. The alternative parametrisation $\fnl(k)\propto\ln k$ for $\fnl^0=0$,  i.e.~Eq.~(\ref{eq:fnlk}) does however remain valid. We now make use of Eq.~(\ref{eq:fnl_analytic}) for $\fnl$ to motivate this alternative parametrisation.\\

Our objective is to expand Eq.~(\ref{eq:fnl_analytic}) for $\fnl$ about some pivot scale $k=k_0$, which we choose to be the scale at which $\fnl=-5/3$. Fluctuations on observable scales $10^{3}\Kpc\lesssim k^{-1}\lesssim10^{4}\Mpc$ today correspond roughly to $\Delta N\sim8$ $e$--foldings over which these modes left the Hubble radius during inflation. Over this range of $N$, the inflationary background is a slowly varying function of time. Hence, the integral $\mathcal{I}(t)$ given by Eq.~(\ref{eq:Itq}) may be expanded about $N\approx{\rm ln}\,(k/k_0)=0$. To first order in slow--roll we obtain:
\be
\mathcal{I}_{\rm SR} = {\rm ln}\,(k/k_0)\Big[ 1+\e_{\phi_0}{\rm ln}\,(k/k_0) \Big]\,,
\ee
where $\e_{\phi_0}\equiv-\dot H_0/H_0^2$, and a subscript `$0$' denotes evaluation at the pivot scale $k_0$. Substituting the above result for $\mathcal{I}_{\rm SR}$ into the slow--roll solution for the curvaton, Eq.~(\ref{eq:curvatonSR}), and using Eq.~(\ref{eq:sk}) for $s_k$ yields:
\be
s_k=s_0\left[  \frac{e^{-\bar\eta_{\s_0}(n-2)\mathcal{I}_{\rm SR}}}{ 1+ \frac{n}{2}s_0\left[1-e^{-\bar\eta_{\s_0}(n-2)\mathcal{I}_{\rm SR}} \right]} \right]\,,
\label{eq:skN}
\ee
where $s_0$ is the value of $s_k$ at the pivot scale:
\be
s_0=\frac{2}{n(n-1)}\,,
\label{eq:szero}
\ee
which implies $\eta_{\s_0}=2\bar\eta_{\s_0}$, see Eq.~(\ref{eq:eta}).
Finally, substituting Eq.~(\ref{eq:skN}) into Eq.~(\ref{eq:fnl_analytic}) for $\fnl$ and expanding the result to second order in ${\rm ln}\,(k/k_0)$ we obtain:
\be
\fnl(k)=-\frac53+\frac{5na_n}{8\rdec}\,\eta_{\s_0}\,{\rm ln}\,(k/k_0) 
\Bigg[ 1 - \frac{a_n}{4}(n+1)\left( \eta_{\s_0}-\frac{4}{a_n(n+1)}\e_{\phi_0} \right)\,{\rm ln}\,(k/k_0) \Bigg]\,,
\label{eq:fnlk-analytic}
\ee
where $a_n=(n-2)/(n-1)$, and we have used $\eta_{\s_0}=2\bar\eta_{\s_0}$. The scale dependent part of $\fnl$ vanishes for $n=2$ (where $a_n=0$), which is simply the non--interacting limit of the model. The constant term and the term linear in ${\rm ln}\,(k/k_0)$ correspond exactly to the parametrisation of $\fnl$ we introduced in Eq.~(\ref{eq:fnlk}), with
\be
\fnl^0=-\frac53\,, \qquad \gfnl=\frac{5na_n}{8\rdec}\eta_{\s_0}\,.
\label{eq:gfnl}
\ee
In Fig.~\ref{fig:fNL-k} we compare Eq.~(\ref{eq:fnlk-analytic}) against a fully numerical calculation for two different values of $n$ and $\eta_{\s_0}$. For $\eta_{\s_0}>10^{-2}$, the term quadratic in ${\rm ln}\,(k/k_0)$ becomes important, and the scale dependence becomes non--linear in ${\rm ln}\,k$. 

\FIGURE{
\epsfig{file=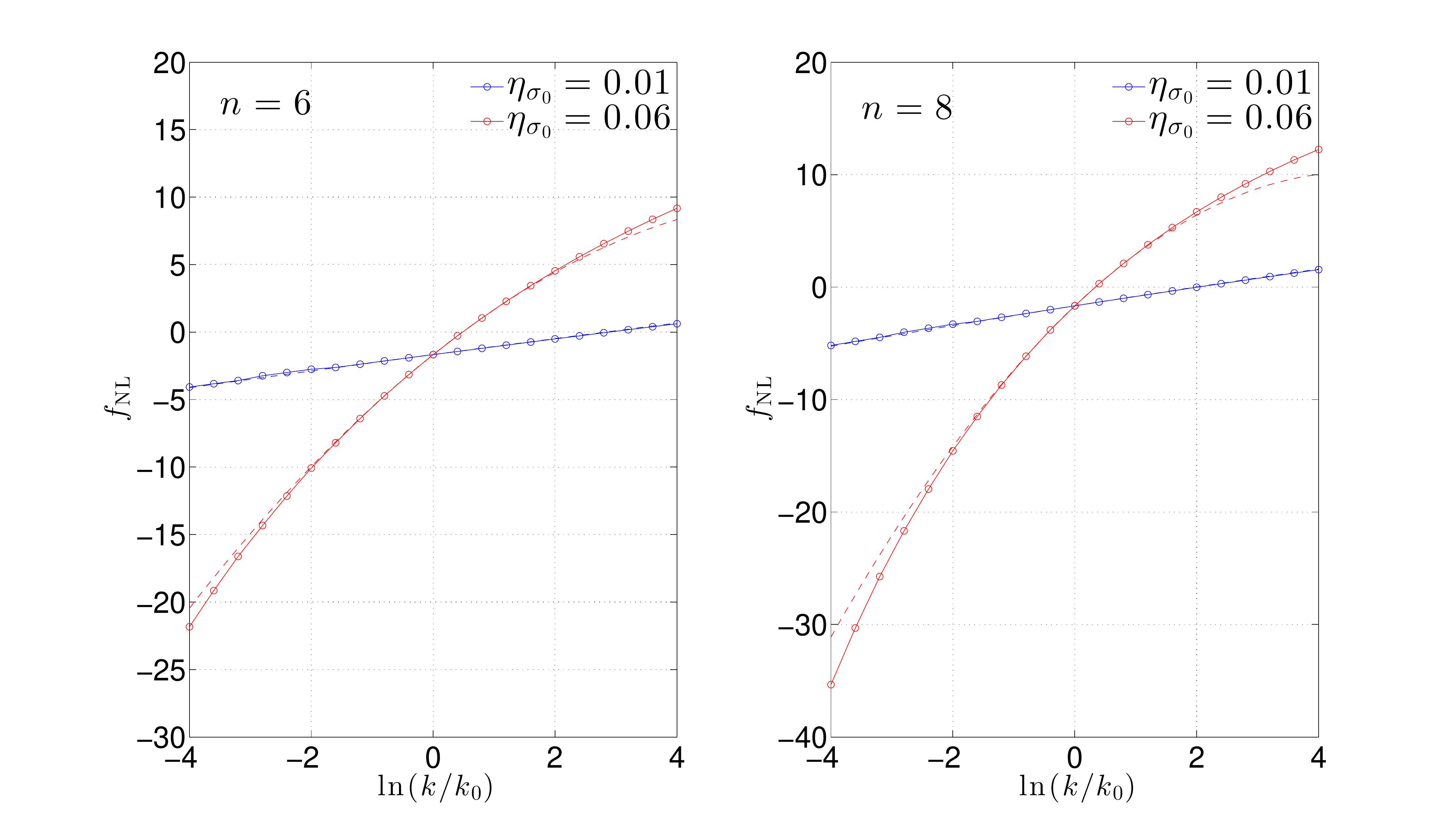,width=16cm} 
\caption{The scale dependence of $\fnl$. Solid lines correspond to a numerical calculation, and dashed lines to our analytic result, Eq.~(\ref{eq:fnlk}). The left panel corresponds to $n=6$ and the right panel to $n=8$. Note the different scales on the $y$--axis. For $\eta_{\s_0}\gtrsim\mathcal{O}(10^{-2})$, $\fnl$ becomes a non--linear function of ${\rm ln}\,(k/k_0)$. For the numerical calculation, we choose $V(\phi)=\mphi\phi^2/2$, $\phi_k=16$. We have chosen an arbitrary value $\rdec=0.05$.}
\label{fig:fNL-k}
}

%-------------------------------------------------------------------------------------------------------------
\section{The CMB Power Asymmetry}\label{sec:asymm}

Both WMAP and Planck have observed a dipole asymmetry in the power spectrum on large scales at about $3\sigma$ significance. This is parametrised by
\bea
P_\zeta&=&P_{\rm iso}\left(1+2 A \hat{\bf n}.\hat{\bf p}\right)\,, 
\eea
where $P_{\rm iso}$ is the isotropic power spectrum, $\hat{\bf n}$ is a direction on the sky of $\zeta$ and $\hat{\bf p}$ is the direction in which the asymmetry is maximised. The current best measurement is $A=0.06-0.07$ \cite{Ade:2015lrj} which is based on measurements on large CMB scales, $l=2-64$, while the smaller scales show no evidence for an asymmetry, requiring $A<0.0045$ for $l=601-2048$ \cite{Flender:2013jja,Adhikari:2014mua,Quartin:2014yaa}. Explaining the asymmetry from an inflationary model requires both a large amplitude mode with wavelength larger than the horizon scale today and non--Gaussianity to correlate this very large scale mode with the CMB scale modes which are asymmetric~\cite{Kanno:2013ohv,Abolhasani:2013vaa,Lyth:2014mga}. The super horizon scale mode must originally be an isocurvature perturbation and cannot be generated from single--field inflation \cite{Erickcek:2008sm,Lyth:2013vha}. It's existence is typically just postulated, but might be related to the beginning of inflation~\cite{Mazumdar:2013yta} or global curvature \cite{Liddle:2013czu}. The asymmetry could also be associated to a domain wall present during inflation \cite{Jazayeri:2014nya}. Because the asymmetry is not observed on small scales, both $A$ and $\fnl$ are required to be strongly scale dependent, which if parametrised by a power law would need to scale like \cite{McDonald:2014kia}
\be
A\propto \fnl \propto k^{\nfnl}\,, \qquad \nfnl<-0.56\,.
\ee 
Such a large scale dependence normally invalidates the assumptions going into the usual calculation of $\nfnl$, Eq.~(\ref{eq:nfnl_std}), if applied over more than about 2--4 $e$--foldings, making this parametrisation inadequate for studying the power asymmetry. For example, one expects ${\rm d}\nfnl/{\rm d}\ln k\sim \nfnl^2$ when $\nfnl$ is larger than the slow--roll parameters \cite{Byrnes:2010xd,Huang:2010cy,Byrnes:2011gh}, in which case $\nfnl$ does not remain constant over much more than one $e$--folding when $\nfnl\sim1$. 

There is a strong tension between the requirement for a sufficiently large non--Gaussianity  to explain the asymmetry and the stringent Planck constraints on both the bispectrum and the low $l$ multipoles  \cite{Namjoo:2013fka,Wang:2013lda,Kanno:2013ohv,Lyth:2014mga,Namjoo:2014nra,Kobayashi:2015qma}, although a joint analysis of scale dependent $\fnl$ and scale dependent $A$ should really be made to properly quantify the tension. Ignoring the scale dependence, the tension is quantified by the following equation (see e.g.~\cite{Kanno:2013ohv} for the derivation), which relates $A$ and $\fnl$ with the dipolar--induced quadrupole asymmetry, $C_2^{(A)}$:
\be
\sqrt{\frac{C_2^{(A)}}{4.2\times10^{-11}}}\frac{\fnl}{10}=6.6\left(\frac{A}{0.07}\right)^2, 
\ee
where we have inserted the observed value of the squared quadrupole temperature anisotropy. While this does not have to correspond to the quadrupole generated by $A$, it can only be smaller if there is a finely tuned cancellation between the intrinsic quadrupole and this additional contribution. 

We are therefore motivated to study scenarios in which $\fnl$ is large (order of 100~\cite{Kanno:2013ohv}) on large scales $l\lesssim64$ but reduces by an order of magnitude on small scales $l\gtrsim700$, which may be easier to arrange by making $\fnl$ cross zero on small scales.  This corresponds to roughly $\gfnl\sim5$. The Planck forecast to be able to measure an error bar on $\nfnl$ of \cite{Sefusatti:2009xu}
\be 
\sigma_{\nfnl}\simeq0.1 \frac{50}{\fnl}\,, 
\ee
suggests $\gfnl\sim5$ may be detectable by Planck. This forecast was made assuming a fiducial value of $\fnl=50$ and constant $\nfnl$ however, and so the analysis needs to be remade with the new parametrisation and new observational constraints. In order to do this, the full shape dependence of the bispectrum is required. \\

We denote the super horizon mode (really the variation of $\sigma$ over the scale of the horizon) by $\Delta\sigma$, which is larger than a ``normal" mode by an enhancement factor $E$, i.e.~$\Delta\sigma=  E H_k/(2\pi)$. This long wavelength mode will act as a shift in the classical background value of the curvaton when the modes inside our horizon exit, meaning that opposite sides of today's horizon will be sensitive to $N_\sigma$ for different values of $\sigma_k$. This concept was described in the context of inhomogeneous non--Gaussianity in \cite{Byrnes:2011ri,Nelson:2012sb,Nurmi:2013xv,LoVerde:2013xka,Baytas:2015nja}. Using a $\delta N$ expansion, and assuming a single--source model for the perturbations (meaning that only one field, such as a curvaton generates the primordial perturbation), we have up to second order in $\Delta\sigma$
\bea
\label{N-Delta-sigma}  N_\sigma(\sigma+\Delta\sigma)=N_\sigma+N_{\sigma\sigma}\Delta\sigma+\frac12N_{\sigma\sigma\sigma}\Delta\sigma^2\,, \\
\label{DeltaP:single} 2A=\frac{\Delta P_\zeta}{P_\zeta}=\frac{N_\sigma(\sigma+\Delta\sigma)^2-N_\sigma^2}{N_\sigma^2}=\frac{12}{5} \fnl N_\sigma\Delta\sigma\simeq 2 \fnl E \sqrt{{\cal P}_\zeta}\simeq \fnl E 10^{-4}\,, \label{eq:AE} 
\eea
where 
\bea
\label{P:single} N_\sigma=\frac{\partial N}{\partial \sigma_k}\,, \qquad
 {\cal P}_\zeta=N_\sigma^2\left(\frac{H_k}{2\pi}\right)^2\,, \qquad \fnl=\frac56\frac{N_{\sigma\sigma}}{N_{\sigma}^2}\,. 
\eea

Requiring $|\fnl|\lesssim10^2$, which is reasonable since we should consider the constraint on $\fnl$ from large scales alone, we require $E\gtrsim 10$, i.e.~a super horizon mode whose amplitude is about 10 times larger than the typical sub horizon scale mode. The variation of the power spectrum due to long wavelength modes is measured by the squeezed limit of the bispectrum, which is closely related to local $\fnl$. From (\ref{DeltaP:single}) we can see that $\fnl\sim10^3$ would naturally lead to $A\sim0.1$, but the observational upper bound on $\fnl$ means that we instead need an unexpectedly large super horizon perturbation, $E\gtrsim10$, to generate a sufficiently large modulation of the power spectrum.

%-------------------------------------------------------------------------------------------------------------
\subsection{The large trispectrum and quadrupolar power asymmetry}\label{sec:g}

When finding a regime with large and scale dependent $\fnl$, we must also check whether $\gnl$ is within observational bounds. From Eq.~(\ref{eq:curvaton_fnl}), we have $\gnl=25g_{\rm osc}/(24\rdec^2)$ to leading order in $1/r_{\rm dec}$.  We are interested in regimes where $|f_{\rm osc}|\ll1$, see (Eq.~\ref{eq:curvaton_fnl}), but these do not correspond to regions where $\gnl$ is accidentally suppressed so we should expect $\gnl\gg\fnl^2$, see e.g. Fig.~3 of \cite{Byrnes:2011gh}. It is straightforward to extend the analytic approximation of Sec.~\ref{sec:analytic_sol} to $\gnl$. To first order in ${\rm ln}\,(k/k_0)$, we find in the limit $\rdec\ll1$ and $s_k\ll1$:
\be
\gnl(k) = -\frac{25}{24\rdec^2}na_n\left[ 1-\frac{n}{2}\frac{(n-3)}{(n-1)}\,\eta_{\s_0}\,{\rm ln}\,(k/k_0)  \right]\,.
\label{eq:gnl_analytic}
\ee
Here, as before, $k_0$ corresponds to the scale at which $\fnl=-5/3$, and $a_n=(n-2)/(n-1)$.  For $\eta_{\s_0}\lesssim0.1$, we see that over the observable range of scales $\gnl$ is approximately scale invariant:  $\gnl\sim -n/\rdec^2$. Using Eq.~(\ref{eq:gfnl}) to substitute for $\rdec$ we find
\be
|\gnl|\sim 4\times10^5\left(\frac{6}{n}\right) \left( \frac{\gfnl}{10} \right)^2 \left(  \frac{10^{-2}}{\eta_{\s_0}} \right)^2.
\label{eq:gnlgfnl}
\ee
This almost saturates the current observational bounds on $\gnl$, which are roughly $|\gnl|\lesssim 5\times10^5$ at $2$--$\sigma$ \cite{Regan:2013jua,Giannantonio:2013uqa,Leistedt:2014zqa,Feng:2015pva,Smith:2015uia} and recently tightened by more than a factor of two with the 2015 Planck results \cite{Ade:2015ava}. The problem can be alleviated for larger values of $\eta_{\s_0}$, however this will typically cause tension with the spectral index measured at the pivot scale $k_0$, which for the self--interacting curvaton scenario is given by
\be
\nz^0-1=-2\e_{\phi_0}+2\eta_{\s_0}\,. \label{ns}
\ee
Furthermore, the term quadratic in ${\rm ln}\,k$ becomes relevant for large $\eta_{\s_0}$ (see Eq.~(\ref{eq:fnlk-analytic})), at which point Eq.~(\ref{eq:gnlgfnl}) breaks down. It is possible to generate a strongly scale--dependent $\fnl$ yet keep $\eta_{\s_0}\sim10^{-2}$ if one considers $s_k>1$. In this regime, the spectral index can be kept compatible with observational constraints, however our numerical investigations confirm that an unacceptably large $\gnl$ is still generated. As an example, we show in Fig.~\ref{fig:fnlgnl} a realisation of the self--interacting curvaton model with $s_k>1$ in which $\fnl$ varies linearly from 30 on the largest scales to $-10$ on the smallest scales, crossing zero in the middle of the range of scales where the asymmetry is not observed.  Without a dedicated analysis it is unknown whether such an $\fnl$ is compatible with the Planck constraint on constant $|\fnl|\lesssim10$, but the figure of $\fnl$ as function of $l_{\rm max}$ from Planck13 suggests it may be \cite{Ade:2013ydc}. The scale dependence given in this example corresponds to $\gfnl\simeq5$, which we can accurately treat as being close to constant over the observed range of scales provided that max$[\epsilon_\phi,\eta_\sigma]\ll 0.1$, see Eq.~(\ref{eq:fnlk}). 
\FIGURE{
\epsfig{file=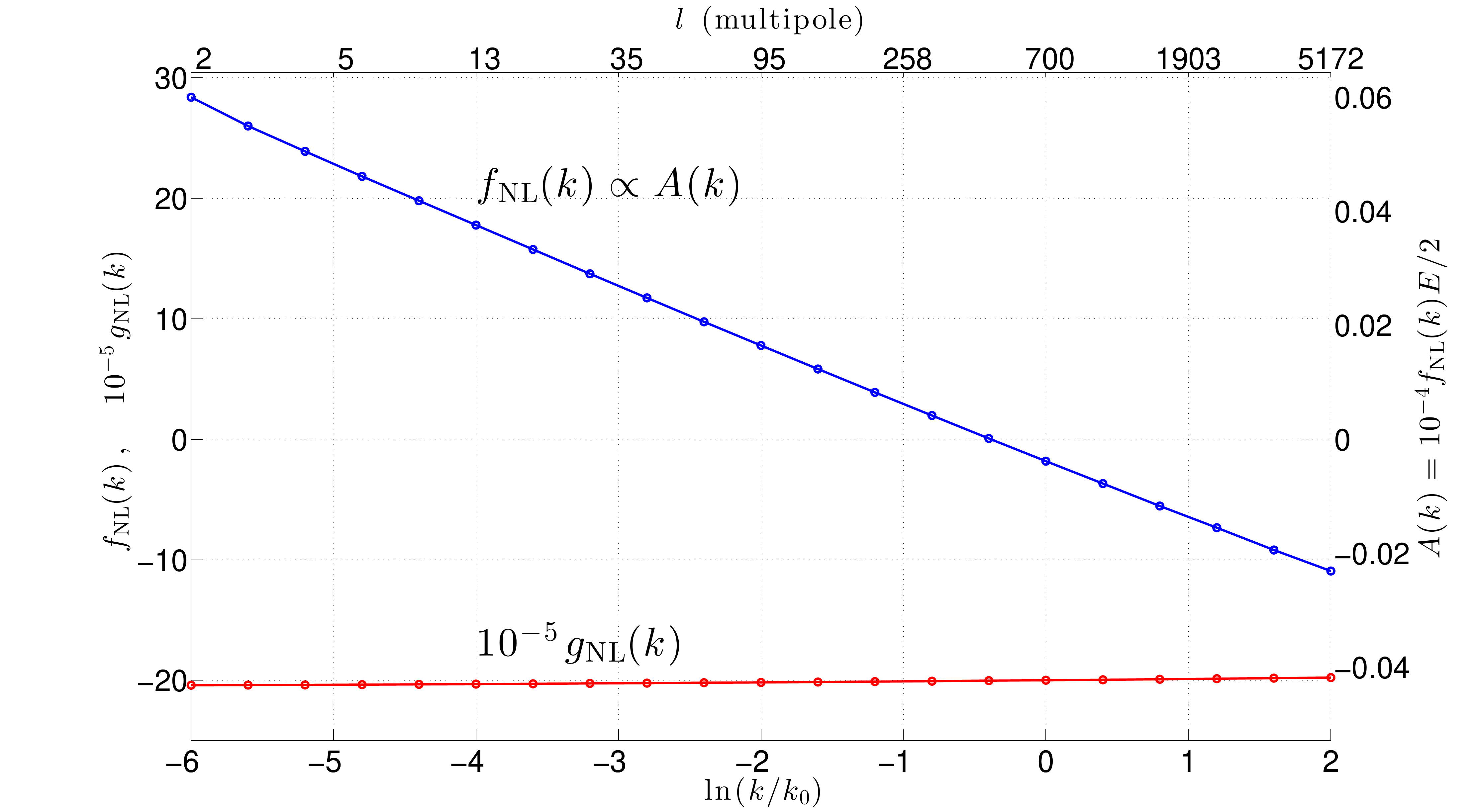,width=16cm} 
\caption{The scale dependence of $\fnl\propto A$ (blue line) and $\gnl$ (red line) for $n=6$, $s_0\simeq3.85$, $\eta_{\s_0}=10^{-2}$ and $\rdec=0.05$. The left--hand $y$--axis corresponds to the values of $\fnl$ and $10^{-5}\gnl$, and the right--hand $y$--axis to the corresponding values of the asymmetry parameter $A=10^{-4}\fnl E/2$, for $E=42$. Whilst the scale dependence of $\fnl$ is able to explain the CMB power asymmetry, the large value of $\gnl\simeq-2\times10^6$ is ruled out. This large value of $\gnl$ induces an unacceptably large quadrupolar asymmetry of the power spectrum: From Eq.~(\ref{eq:g}) we find $B\sim-14$, which is almost four order of magnitude larger than the constraint $B=0.002\pm0.016$ of Ref.~\cite{Kim:2013gka}. In this figure, $\fnl$ and $\gnl$ were computed numerically, with $V(\phi)=\mphi\phi^2/2$, $\phi_k=16$. To convert between wave number $k$ and multipole $l$ we have used the approximate relation $l\approx x_{\rm lss}k$, where $x_{\rm lss}\approx1.4\times10^4\Mpc$ is the distance to the surface of last scattering. The pivot scale has been chosen to match the Planck pivot scale $k_0=0.05\Mpc^{-1}$.}
\label{fig:fnlgnl}
}
The observed value of the spectral index implies that this inequality should be satisfied and hence $\fnl$ will be linear in ${\rm ln}\,k$. Notice that a constant $\gfnl$ is incompatible with a scale dependent $\fnl$ and constant $\nfnl$, given the relation $\gfnl\simeq\fnl\nfnl$, showing that our new parametrisation is much more accurate in cases where $\nfnl\sim1$. We also remark that there may be a large running of the spectral index due to the scale dependence of $\eta_\s$ (a correlation between this and large $\gnl$ has been recently explored in \cite{Enqvist:2014nsa,Biagetti:2015tja}), but this can be alleviated by making the contribution of $\eta_\s$ to the spectral index in (\ref{ns}) subdominant. We have numerically checked that the running is much smaller than the observational bounds in all of our explicit examples.

A large $\gnl$ is associated with a large induced quadrupolar asymmetry of the power spectrum. We parametrise this by $B$,
\bea 
P_\zeta&=&P_{\rm iso}\left(1+2 A \hat{\bf n}.\hat{\bf p}+B\left( \hat{\bf n}.\hat{\bf p}\right)^2\right)\,,  \label{Pexpand}
\eea
and is given by
\bea B=\left[\frac{54}{25}\gnl+\left(\frac{6}{5}\fnl\right)^2\right](N_\sigma\Delta\sigma)^2\simeq 2\gnl E^2 {\cal P}_\zeta = 4 \gnl E^2 10^{-9}\,,  \label{eq:g}
\eea
where
\bea 
\frac{54}{25}|\gnl|=\Big| \frac{N_{\sigma\sigma\sigma}}{N_\sigma^3}\Big| \gg \left(\frac{6}{5}\fnl\right)^2\,. 
\eea
No direct observational constraint on $B$ has been made, but an essentially equivalent quantity was constrained in the Planck2013 analysis of non--Gaussianity, see Fig.~17 of \cite{Ade:2013ydc}, where the data point with $L=1$ corresponds to a rescaled version of $A$ and the data point with $L=2$ corresponds to $B$ (where both parameters are assumed to be constant in the data analysis)\footnote{We thank Zac Kenton and Antony Lewis for a discussion about this point. A recent paper by Kenton {\it et al} considers the possibility that asymmetry is generated by the next (third) order term which would appear in (\ref{Pexpand}) \cite{Kenton:2015jga}. The notation in that paper is related to ours by $A=A_1$ and $B=2A_2$.}. This analysis did not require the preferred direction associated to $A$ and $B$ to be that same, unlike the prediction of our model but doing so should only make the constraint tighter. As an order of magnitude estimate, we require $|B|\lesssim10^{-2}$ but none of our conclusions are sensitive to the exact value\footnote{A comparable constraint was found on the related parameter $g_*$ defined in Fourier space \be
P=P_{\rm iso}\left(1+g_* (\hat{\bf n}.\hat{\bf p})^2\right)\,.
\ee
on which Kim and Komatsu \cite{Kim:2013gka} found the constraint $g_*=0.002\pm0.016$, (see also \cite{Ade:2015oja} for the comparable Planck 2015 results). Their analysis did not require the direction of the quadrupolar power asymmetry to correspond to direction of the dipolar power asymmetry. Despite being defined in Fourier space, a preferred direction in Fourier space will generate a preferred direction on the CMB due to our view of it being a projection on a sphere.} .
We note that there could be an independent contribution to $B$ if there is more than one large, super horizon scale mode but one would have to tune both the amplitude and direction of the second mode in order to arrange a cancellation to the term we calculated here. For the example given in Fig.~\ref{fig:fnlgnl}, where $E\sim40$ and $|\gnl|\sim10^6$ we find $|B|\sim14$, which is about four orders of magnitude too large. While we cannot conclude that this is a problem for all models, and there are many free parameters to play with, we caution that parameter regions with large and strongly scale dependent $\fnl$ may be associated with too large values of $\gnl$ and $B$, and their values should be checked whenever constructing an explicit model. A relation between large $\gnl$ and regions of scale dependent $\fnl$ was also spotted in \cite{Enqvist:2014nsa} in the context of models with an observable running of the spectral index, but as far as we are aware the association with large $B$ has not been made before.

Another way to see why $B$ is typically too large is remove the $E$ dependence from Eq.~(\ref{eq:g}) using Eq.~(\ref{eq:AE}),
\be 
B\simeq 10^{-2} \frac{\gnl}{\fnl(k)^2}\left(\frac{A(k)}{0.07}\right)^2\,, 
\label{eq:gstar2}
\ee
where care should be taken to evaluate $A$ and $\fnl$ at the same pivot scale, since these terms are strongly scale dependent but $B$ is not. Even for large self--interaction strengths, where no analytic solution exists, we see from Eq.~(\ref{eq:curvaton_fnl}) that regions where $\fnl$ crosses zero quickly (i.e.~where $\gfnl$ is large) will have $\rdec\ll1$ and $|f_{\rm osc}|\ll1$. Such parameter regions do not correspond to an accidentally suppressed $g_{\rm osc}$, so we generically expect $|B|\gg0.01$ in the self--interacting curvaton scenario. It would be interesting to see if this conclusion applies to other models generating local non--Gaussianity.  We remark that Eq.~(\ref{eq:gstar2}) applies to all single--source models (those in which perturbations from only one field generate the primordial curvature perturbation)  and is not restricted to the self--interacting curvaton.

Our parametrisation of $\fnl$ growing logarithmically is related to models in which a large ``loop" generates the dominant contribution to $\fnl$, which arises in cases where the tree level term to the bispectrum is suppressed, and such models also tend to have a large trispectrum \cite{Boubekeur:2005fj,Suyama:2008nt,Bramante:2011zr}. However in such cases the role of the cut--off momenta is not fully understood \cite{Seery:2010kh}, while our results are manifestly invariant of the pivot scale (which replaces the cut--off scale). See also the discussion in Sec.~3.2 of \cite{Byrnes:2011gh}, and for a related perspective on the fine tuning required to have $|\gnl|\ggg|\fnl|$ see \cite{Byrnes:2013qjy}.

A curvaton potential which allows $\eta_\sigma<0$ may be preferred due to the observed value of the spectral index, and there has been interest in the axionic curvaton which does also generate a scale--dependent $\fnl$ \cite{Huang:2010cy}. However in this model the scale dependence is closely related to the running of the spectral index, which is tightly constrained by data to be small \cite{Kawasaki:2012gg}. Hence the axionic curvaton cannot have a sufficiently strongly scale--dependent non--Gaussianity to explain the scale dependent power asymmetry.

Overall, we conclude that there is no model of the dipolar power asymmetry which has been demonstrated to meet all observational requirements. The requirement for a strong scale dependence to this modulation means that the usual formalism of constant $\nfnl$ will not be valid over the observed range of scales, while the formalism we introduced in this paper of $\gfnl={\rm d}\ln k/{\rm d} \fnl$ typically remains constant over the observed eight $e$--foldings, even when $|\gfnl|\gg1$. Regardless of the parametrisation used, we have shown that there is a high danger that $\gnl$ and/or $B$ will be too large to match observations and their values should be checked when constructing models.

%-------------------------------------------------------------------------------------------------------------
\section{Conclusions}\label{sec:conc}

In this paper we have introduced a new parametrisation for the scale dependence of $\fnl$, which unlike previous parametrisations remains valid when $\fnl$ changes sign, and when the scale dependence is large. The parametrisation is simply given by
\bea 
\fnl(k)=\fnl^0+\gfnl \ln\left(\frac{k}{k_0}\right)\,,
\label{conc:fnl}
\eea
where 
\bea 
\gfnl=\frac{{\rm d} \fnl}{{\rm d}\ln k}\,.
\eea
We expect $\gfnl$ to be approximately constant, and hence any higher  order terms in Eq.~(\ref{conc:fnl}) to be negligible, over a range which spans $\ln(k/k_0)$ inversely proportional to the slow--roll parameters, which is a much larger range than the seven $e$--foldings that we are able to constrain with CMB and LSS data.

Given the tight observational constraints on constant $\fnl$, our parametrisation is of most interest in the case $\fnl^0\simeq0$, and hence where there is a significant chance of $\fnl$ crossing zero within the observable range of scales. Our formalism is only valid for an equilateral configuration of the bispectrum in Fourier space, where the derivative depends on the single characteristic length scale. In order to observationally constrain our parametrisation, it is necessary to be able to calculate the full bispectral shape. We leave this for future work. 

We have demonstrated the applicability of our formalism to the self--interacting curvaton scenario, a model in which even a small correction to the mass term from self interactions can lead to strongly scale--dependent $\fnl$. We have found the first analytical solution which remains valid if the curvaton evolves during inflation and the inflaton reheating, which despite being restricted to the limit of small self--interactions, is valid within the sufficiently non--linear regime to accurately describe $\fnl$ where it crosses zero. As a by product, we find that such regimes of enhanced scale dependence and $\fnl$ crossing zero do exist even in the limit of a quartic self--interaction term, which had not been previously found due to past analyses neglecting the evolution of the curvaton before the universe becomes radiation dominated after the inflaton has reheated. For arbitrary values of the self--interacting strength, we have found (\ref{conc:fnl}) to be a good match to numerical solutions, which also match the analytical solutions in the appropriate limit. 

We have applied our formalism to the observed power asymmetry of the primordial power spectrum. The asymmetry is required to be strongly scale dependent, since it is detected to be about a $10\%$ effect on large scales $l<62$, but must be at least an order of magnitude smaller on small scales, $l>600$. This requires a similarly large decrease in $\fnl$, something which our parametrisation can cope with but a commonly used previous one of constant $\nfnl={\rm d}\ln|\fnl|/{\rm d}\ln k$ can not. The reason is that the previous formalism required $|\nfnl \ln(k/k_0)|\lesssim1$, while observations require $\nfnl\lesssim-0.5$. 

While we can easily fine--tune curvaton models to have a sufficiently strong scale-dependent $\fnl$, we find there are two generic new challenges which such a model must face. In parameter regions where  $\fnl$ is close to crossing zero and strongly scale dependent, we find that both $\gnl$ and the quadrupolar power asymmetry are generically large and close to scale invariant. For the many examples we have tested,  $B$ and typically also $\gnl$ are too large to agree with observations. In addition, there is the standard problem of explaining the origin of a very large amplitude super horizon scale perturbation which the power asymmetry requires. 

%-------------------------------------------------------------------------------------------------------------
\acknowledgments{The authors thank Zac Kenton for correspondence relating to our constraint on the quadrupolar power asymmetry $B$, for which we wrongly directly applied 
the constraint on $g_*$ in arXiv v1 of this paper. We also thank Takeshi Koboyashi, Antony Lewis, Donough Regan, Tomo Takahashi and especially David Seery for valuable comments on a draft of this paper, and the Centro de Ciencias de Benasque Pedro Pascual for hospitality during a workshop on Modern Cosmology during August 2014 where this project began. CB is supported by a Royal Society University Research Fellowship. The research leading to these results has received funding from the European Research Council under the European Union's Seventh Framework Programme (FP/2007--2013) / ERC Grant Agreement No. [308082]. }
%-------------------------------------------------------------------------------------------------------------

%-------------------------------------------------------------------------------------------------------------
\appendix
%-------------------------------------------------------------------------------------------------------------
\section{General Expression for $\gfnl$}\label{sec:app:gfnl}

Here we provide the general $\delta$N expression for the bispectrum running parameter $\gfnl$. A sum over $I,J,K,...$ is implicitly assumed. We start with the full expression for $\fnl$:
\be
\fnl =\frac56 \frac{N_{IJ}N_IN_J}{( N_KN_K)^2}\,, 
\label{eq:app:fnl}
\ee
where $N_I\equiv\partial N/\partial\vp_{I_k}$, etc and subscript $k$ denotes evaluation of a quantity at the epoch of Hubble exit of mode $k$. Assuming slow--roll at horizon--crossing for all fields $\vp_I$ we can write 
\be
\frac{\rm d}{{\rm d\,ln}\,k}\approx \frac{\dtvp_{I_k}}{H_k}\frac{\partial}{\partial\vp_{I_k}}\,,
\label{eq:app:dk}
\ee
where the sum is carried out over all fields $\vp_I$. This enables us differentiate Eq.~(\ref{eq:app:fnl}) with respect to $k$:
\bea
\gfnl &\equiv& \frac{{\rm d}\,\fnl}{{\rm d\,ln}\,k}\Bigg|_{k=k_0} \nonumber \\
         &=& -2\fnl^0(n_{\zeta}^0 - 1 +2\e_0)+\left(\frac{5}{6H_0}\right)\left[\frac{N_{IJK}N_{I}N_{J}\dtvp_{K_0}}{(N_LN_L)^2} 
+ 2\frac{N_{IJ}N_{IK}N_{J}\dtvp_{K_0}}{(N_LN_L)^2}\right]\,,
\label{eq:app:gfnl}
\eea
where the spectral index reads:
\be
n_{\zeta}^0-1=-2\e_0+\frac{2}{H_0}\frac{ \dtvp_{J_0}N_{IJ}N_I }{N_KN_K}\,,
\label{eq:app:nz}
\ee
and a superscript/subscript $0$ denotes evaluation at the pivot scale $k_0$. Using $\frac{{\rm d}N}{{\rm d}t_k}=-H_k$ and the slow--roll field equations, one can show that
\bea
N_{I}W_{I_0} &=& W_{0}  \label{eq:Dn_pot_1st}\,, \\
N_{IJ}W_{I_0} &=& W_{J_0} - N_{I}W_{IJ_0}\,,  \label{eq:Dn_pot_2nd} \\
N_{IJK}W_{I_0} &=& W_{JK_0} - N_{IJ}W_{IK_0} - N_{IK}W_{IJ_0} - N_{I}W_{IJK_0} \,, \label{eq:Dn_pot_3rd}
\eea 
where Eqs.~(\ref{eq:Dn_pot_2nd}) and~(\ref{eq:Dn_pot_3rd}) are derived by differentiating Eq.~(\ref{eq:Dn_pot_1st}) with respect to $\vp_{I_0}$. These expressions allow us replace third derivatives of $N$ with second derivatives yielding the final result
\bea
\gfnl = &-& 2\fnl^0(\nz^0 - 1 +2\e_0)  \nonumber \\  
&-&\frac{5}{6(N_LN_L)^2}\Big[2-4\eta_{IK_0}N_{IJ}N_{J}N_{K} - \eta_{IJ_0}N_{I}N_{J} - \xi_{IJK_0}N_{I}N_{J}N_{K}\Big]\,,
\label{eq:nfnl_pot} 
\eea
where $\eta_{IJ}=W_{IJ}/W$ and $\xi_{IJK}=W_{IJK}/W$. In the limit where the primordial perturbations are generated by a curvaton field, $\s$,  (the single source limit) the expression for $\gfnl$ simplifies to:
\be
\gfnl = \frac{W^0_{\s\s\s}}{3H_0^2}\frac{5}{6N_\s}\simeq \frac{W^0_{\s\s\s}}{3H_0^2} 0.3 r^{1/2} \qquad \qquad {\rm (single\,\,source)}\,,
\ee
where $r\simeq8/N_\s^2$ is the tensor-to-scalar ratio. Although $r$ must be small,  $W^0_{\s\s\s}/H_0^2$ is not a slow--roll quantity and it may be much larger than unity in models with a non--quadratic potential. This result is obtained by carefully taking the limit 
\be
\sqrt{\frac{\e_{J_0}}{\e_{\s_0}}}\frac{N_J}{N_\s}\ll 1\,, \qquad 
\frac{\eta_{JJ_0}}{\eta_{\s\s_0}}\left(\frac{N_J}{N_\s}\right)^2\ll 1 \qquad (J\neq\s)\,,
\ee
in Eq.~(\ref{eq:nfnl_pot}). One can easily verify that inserting $N_\s=(2/3)\rdec(\sosc^\prime/\sosc)$ with $\sosc$ given by Eq.~(\ref{eq:curvatonSR2}) into the above equation, one obtains the result for $\gfnl$ given in Eq.~(\ref{eq:gfnl}), as expected.

%-------------------------------------------------------------------------------------------------------------
%\bibliographystyle{jhep}
\bibliographystyle{JHEP}
\bibliography{bib_file}

\end{document}